\documentstyle[12pt]{article}
\input{psfig}
\input{epsf}

\def\simge{\mathrel{%
   \rlap{\raise 0.511ex \hbox{$>$}}{\lower 0.511ex \hbox{$\sim$}}}}
\def\simle{\mathrel{
   \rlap{\raise 0.511ex \hbox{$<$}}{\lower 0.511ex \hbox{$\sim$}}}}
 
\def\slashchar#1{\setbox0=\hbox{$#1$}           
   \dimen0=\wd0                                 
   \setbox1=\hbox{/} \dimen1=\wd1               
   \ifdim\dimen0>\dimen1                        
      \rlap{\hbox to \dimen0{\hfil/\hfil}}      
      #1                                        
   \else                                        
      \rlap{\hbox to \dimen1{\hfil$#1$\hfil}}   
      /                                         
   \fi}                                         %
\def\nn{\nonumber}
\def\ts{\thinspace}

\def\ra{\rightarrow}

\def\ol{\bar}

\def\be{\begin{equation}} 
\def\ee{\end{equation}} 
\def\bea{\begin{eqnarray}}
\def\eea{\end{eqnarray}}
\def\ba{\begin{array}}
\def\ea{\end{array}}
\def\dag{\dagger}

\def\CH{{\cal H}}

\def\CM{{\cal M}}

\def\CO{{\cal O}}

\def\getc{g_{ETC}}

\def\kslash{\raise.15ex\hbox{/}\kern-.57em k}

\def\METC{M_{ETC}}
\def\tpi{\pi_T}

\def\gev{{\rm GeV}}
\def\tev{{\rm TeV}}

\def\half{{\textstyle{ { 1\over { 2 } }}}}

\begin{document}
\title{
\vskip -15mm
\begin{flushright}
\vskip -15mm
{\small BUHEP-00-1\\
Fermilab-Pub-00/008-T\\
hep-ph/0001056\\}
\vskip 5mm
\end{flushright}
{\Large{\bf \hskip 0.38truein
Vacuum~Alignment~in~Technicolor~Theories\\ I. The Technifermion Sector}}\\
}
\author{
\centerline{{\small Kenneth Lane$^{1}$\thanks{lane@buphyc.bu.edu}\ts\ts,}
{\small Tongu\c c Rador$^{1}$\thanks{rador@budoe.bu.edu}\ts\ts, and}
{\small Estia Eichten$^{2}$\thanks{eichten@fnal.gov}}}\\
\centerline{{\small {$^{1}$}Department of Physics, Boston University,}}\\
\centerline{{\small 590 Commonwealth Avenue, Boston, MA 02215}}\\
\centerline{{\small {$^{2}$}Fermilab, P.O.~Box 500, Batavia, IL 60510}}\\
}
\maketitle
\begin{abstract}
We have carried out numerical studies of vacuum alignment in technicolor
models of electroweak and flavor symmetry breaking. The goal is to understand
alignment's implications for strong and weak CP nonconservation in quark
interactions. In this first part, we restrict our attention to the
technifermion sector of simple models. We find several interesting phenomena,
including (1) the possibility that all observable phases in the
technifermions' unitary vacuum--alignment matrix are integer multiples of
$\pi/N'$ where $N' \le N$, the number of technifermion doublets, and (2) the
possibility of exceptionally light pseudoGoldstone technipions.

\end{abstract}


\newpage

\section*{1. Introduction}

One of the original motivations for the dynamical approach to electroweak
{\it and} flavor symmetry breaking---specifically, technicolor~\cite{tc} and
extended technicolor~\cite{etceekl,etcsd}---was the belief that it would
solve the problem of strong CP--violation in QCD~\cite{CPreview}. The idea
was this: In a theory consisting only of gauge interactions of {\it massless}
fermions, instanton angles such as $\theta_{QCD}$ may be freely rotated to
zero. Purely dynamical masses, i.e., fermion bilinear condensates, may be
assumed to be CP--conserving. And, the fermions' hard masses are generated by
the joint action of dynamical and explicit chiral symmetry breaking, all
induced by gauge interactions {\it alone}. It was hoped that this combination
naturally would produce a for which $\ol \theta_q = \theta_{QCD} +
\arg\ts\det (M_q) = 0$ without an axion. This is naive, especially if at
least some of the observed CP--violation is to emerge from diagonalizing the
quark mass matrix $M_q$.

In fact, the way to determine the true status of CP symmetry in a
superficially CP--invariant theory was prescribed long ago by
Dashen~\cite{rfd}. He studied the question of determining the correct
perturbative ground state $\vert \Omega \rangle$ upon which to begin an
expansion about the chiral limit. This process is known as vacuum alignment.
When the chiral symmetry of quarks is spontaneously broken, there are
infinitely many degenerate vacua, parameterized by transformations
corresponding to massless Goldstone bosons. Dashen showed that, if this
chiral symmetry is also explicitly broken by $\CH'_q = \ol q_L M_q \ts q_R +
{\rm h.c.}$, the degeneracy is lifted and the correctly aligned zeroth--order
ground state $\vert \Omega \rangle$ is the one in which the expected value of
$\CH'_q$ is least. In practice, it is easier to fix $\vert \Omega \rangle$ as
a ``standard vacuum'' with simple condensates~\footnote{We assume that the
quark chiral symmetry $SU(2n)_L \otimes SU(2n)_R$ is spontaneously broken to
an $SU(2n)$ subgroup, in which case the quark condensates $\langle\ol q_{aL}
q_{bR} \rangle$ are proportional to an $SU(2n)$ matrix. In the standard
vacuum, $\langle \Omega|\ol q_{aL} q_{bR} |\Omega\rangle \propto
\delta_{ab}$.} and chirally rotate $\CH'_q$ to find the minimum vacuum
energy. Dashen showed that, even if the original $\CH'_q$ is CP--conserving,
i.e., if $M_q$ is real, the Hamiltonian aligned with $\vert \Omega \rangle$
may be CP--violating. This is spontaneous CP--violation. For real $M_q$, it
occurs if $\ol \theta_q = \pi$. The aligned Hamiltonian has the CP--violating
term $i \nu_q \ts \ol q \gamma_5 q$, where $\nu_q$ is of order the smallest
eigenvalue of $M_q$~\cite{nuyts}.

Dashen's study was made in the context of QCD, but it applies to a theory in
which QCD is united with technicolor to generate quark masses by extended
technicolor~\cite{elp}. In such a theory, the chiral symmetries of
technifermions are spontaneously broken at $\Lambda_{TC} \sim 1\,\tev$,
giving rise to massless technipions, $\tpi$. All but the three $\tpi^{\pm,0}$
that become the longitudinal components of the $W^\pm$ and $Z^0$ bosons must
get large masses, at least 50--70~GeV for the charged ones. Quark chiral
symmetries are spontaneously broken at the much lower scale $\Lambda_{QCD}
\sim 1\,\gev$.~\footnote{Complications due to the top quark will be discussed
below.} All these symmetries, except electroweak $SU(2) \otimes U(1)$,
are explicitly broken by ETC--boson exchange interactions. They are
well--approximated at 1~TeV by four--fermion interactions, $\ol T T \ol T T$
and $\ol q T \ol T q$, suppressed by the square of $\METC \simge
100\,\tev$.~\footnote{To a lesser extent, the electroweak interactions also
contribute to explicit symmetry breaking; see Ref.~\cite{elp}. They are
ignored in Eq.~(2) below.}

It is natural to assume that ETC breaking is such that these four--fermion
interactions have real coefficients and so are superficially
CP--conserving. Vacuum alignment then has three possible outcomes: (1)~the
correct chiral--breaking perturbation, $\CH'$, is still CP--conserving and,
in particular, the Cabibbo--Kobayashi--Maskawa (CKM) matrix is real;
(2)~$\CH'$ is CP--violating, but $|\nu_q| \sim m_u$ is $10^9$ times too
large; (3)~$\CH'$ is CP--violating, but $|\nu_q| = 0$ or is at most is of
order the ratio of condensates $\langle \ol q q \rangle/\langle \ol T T
\rangle \simle 10^{-9}$. This last alternative, of course, is the desired
one. Unfortunately, no physical criteria were found to lead to models of
type~3.

The matter rested there until the dynamical attempts known as
topcolor--assisted technicolor (TC2) were made to deal with the large mass of
the top quark~\cite{tctwohill,tctwoklee}. It has always been difficult for
the dynamical approach, especially extended technicolor, to account for the
top quark's mass. Either the ETC scale generating the top mass must be near
1~TeV, leading to conflict with experimental measurements on the $\rho$
parameter~\cite{iso} and the $Z \ra \ol b b$ decay rate~\cite{zbbth}, or,
if it is made much higher, the coupling $\getc$ must be unnaturally
fine--tuned. Hill circumvented these difficulties by invoking another strong
interaction near 1~TeV, topcolor, to generate a large $\ol t t$ condensate
and top mass. In TC2, ordinary technicolor remains responsible for the bulk
of electroweak symmetry breaking.

An important consequence of this scenario---and this is where vacuum
alignment comes back in---is that top condensation implies a triplet of
massless Goldstone ``top--pions'', $\pi_t^{\pm,0}$. These must acquire mass
$M_{\pi_t} \simge m_t = 175\,\gev$; otherwise $t \ra b \pi_t^+$ becomes a
major decay mode. Extended technicolor interactions provide this mass by
contributing 5--10~GeV to $m_t$~\cite{tctwohill}. At the same time, this ETC
contribution must not induce appreciable mixing of top--pions with ordinary
technipions~\cite{cthill}. Some technipions may be as light as
100~GeV~\cite{elw}, so that large mixing would lead to substantial, and also
unobserved, $t \ra b \pi_T^+$.

Balaji studied top--pion mass and mixing in a specific model, and he obtained
encouraging results~\cite{bbvac}. However, his conclusions are preliminary
because he was unable to execute vacuum alignment properly. This is
understandable because vacuum alignment in TC2 models is very
complicated. Now it involves at least two gauge interactions strong near
1~TeV---technicolor and topcolor---with some technifermions transforming
under both. And, many technifermions are needed to accommodate various
experimental constraints, making the chiral flavor group quite large; see
Ref.~\cite{tctwoklee} for details. One of these experimental constraints is
that no physical technipion be massless or very light. The criterion used in
Refs.~\cite{tctwoklee,bbvac} for deciding this was that no spontaneously
broken chiral charge (other than the electroweak charges) can commute with
the ETC--generated $\ol T T \ol T T$ interactions. We shall see in Section~3
that this criterion, which works in QCD, is {\it insufficient} to guarantee
that all technipions are massive.

The problem of vacuum alignment in technicolor theories is too complex for
analytical treatment. Numerical methods are needed. We start the numerical
analysis in this paper by considering the technifermion sector of a simple
ETC model, one in which there are $N$ doublets of a single type of
technifermion that transforms according to the complex fundamental
representation of the technicolor gauge group $SU(N_{TC})$.

The rest of this paper proceeds as follows: In Section~2 we define our
simplified ETC model and present the formalism in first--order chiral
perturbation theory for vacuum alignment and calculating technipion
masses. There we illustrate the unexpected (to us, anyway) fact that chiral
symmetries are not always manifest in the chiral--breaking perturbation
$\CH'$. We present in Section~3 the main results of vacuum alignment in the
technifermion sector. We have found a quite surprising result: the phases in
the technifermions' unitary vacuum--alignment matrix $W_0$ may be integer
multiples of $\pi/N'$ where $N' \le N$. If they are allowed by unitarity,
these ``rational phases'' occur because the terms in $\CH'$ make it
energetically favorable for certain phases to be equal or to differ by $\pi$
and because $W_0$ is unimodular. If unitarity frustrates this alignment of
phases in $W_0$, they are irrational. We shall see that the rational phases
appear as islands in an irrational sea, the boundaries of which are defined
by critical values of the parameters in $\CH'$.  Furthermore, a technipion
becomes massless, a Goldstone boson to first order, at the island shore,
where the ETC parameters become critical. This has the important
phenomenological consequence that an exceptionally light technipion often
accompanies the rational phases because generically chosen parameters are not
far from the critical ones. Thus, some technipions may be even lighter than
we expected~\cite{elw}, a fact which may be welcome and which, in any case,
can be used to help choose among models. We conclude in Section~4 with a
brief look ahead to vacuum alignment and CP--violation in the quark sector.

\section*{2. The Extended Technicolor Model}

To simplify our numerical studies, we consider models in which a single kind
of technifermion interacts with quarks (but no leptons) via ETC
interactions. There are $N$ technifermion doublets $(U_{i\ts L,R}, \ts D_{i
\ts L,R})$, $i = 1,2,\dots,N$, all transforming according to the fundamental
representation of the technicolor gauge group $SU(N_{TC})$. There are $n$
generations of $SU(3)_C$ triplet quarks $(u_{a\ts L,R}, \ts d_{a\ts L,R})$,
$a = 1,2,\dots,n$. The left--handed fermions are electroweak $SU(2)$ doublets
and the right--handed ones are singlets. Here and below, we exhibit only
flavor, not technicolor nor color, indices. Although it is not essential for
our studies, we shall assume that the technicolor gauge coupling runs slowly,
or ``walks'' from the TC to the ETC scale~\cite{wtc}. No provision to give a
realistic top quark mass, such as topcolor--assisted
technicolor~\cite{tctwohill}, will be made in this paper.

The technifermions are ordinary color--singlets, so the chiral flavor
group of our model is $G_f = \left[SU(2N)_{L} \otimes SU(2N)_{R}\right]
\otimes\left[SU(2n)_{L} \otimes SU(2n)_{R}\right]$. We have excluded
anomalous $U_A(1)$'s strongly broken by TC and color instanton effects. When
the TC and QCD couplings reach their required critical values, these
symmetries are spontaneously broken to $S_f = SU(2N) \otimes SU(2n)$. We
shall take this residual symmetry to be the diagonal vectorial one by
adopting as our standard vacuum the state $|\Omega\rangle$ in which the
nonzero fermion bilinear condensates are diagonal:
\bea\label{eq:standard}
\langle \Omega |\ol U_{iL} U_{jR}|\Omega \rangle &=&
\langle \Omega |\ol D_{iL} D_{jR}|\Omega \rangle = -\delta_{ij} \Delta_T
\nn\\
\langle \Omega |\ol u_{aL} u_{bR}|\Omega \rangle &=&
\langle \Omega |\ol d_{aL} d_{bR}|\Omega \rangle = -\delta_{ab} \Delta_q \ts.
\eea
The condensates $\Delta_T \simeq N_{TC} \Lambda^3_{TC}$ and $\Delta_q \simeq
N_C \Lambda^3_{QCD}$ when they are renormalized at their respective strong
interaction scales. Of 
course, $N_C = 3$.

All of the $G_f$ symmetries except for the gauged electroweak $SU(2) \otimes
U(1)$ must be {\it explicitly} broken by extended technicolor
interactions~\cite{etceekl,etcsd}. In the absence of a concrete ETC model, we
write the interactions broken at the scale $M_{ETC}/g_{ETC} = \CO(100\,\tev)$
in the phenomenological four-fermion form~\footnote{In Eq.~(2), we have not
made any assumption about the structure of ETC interactions vis--a--vis the
electroweak ones.}
\bea\label{eq:Hetc}
\CH' &\equiv& \CH'_{TT} + \CH'_{Tq} + \CH'_{qq} \nn\\
&=& \Lambda^{TT}_{ijkl} \ts \ol{T}_{iL}\gamma^{\mu}T_{jL}
\ts \ol{T}_{kR}\gamma_{\mu}T_{lR} \nn 
+ \Lambda^{Tq}_{iabj} \ts \ol{T}_{iL}\gamma^{\mu}q_{aL}
\ts \ol{q}_{bR}\gamma_{\mu}T_{jR} + {\rm h.c.}\\
&+& \Lambda^{qq}_{abcd} \ts \ol{q}_{aL}\gamma_{\mu}q_{bL}
\ts \ol{q}_{cR}\gamma_{\mu}q_{dR} \ts,
\eea
where $T_{i\ts L,R}$ and $q_{a\ts L,R}$ stand for all $2N$ technifermions and
$2n$ quarks, respectively. Here, $M_{ETC}$ is a typical ETC gauge boson mass
and the $\Lambda$ coefficients are $g^2_{ETC}/M^2_{ETC}$ times mixing factors
for these bosons and group theoretical factors. Typically, the $\Lambda$'s
are positive, though some may be negative. In our calculations, we choose the
$\Lambda$'s to avoid unwanted Goldstone bosons. Hermiticity of $\CH'$
requires
\be\label{eq:herm}
(\Lambda^{TT}_{ijkl})^* = \Lambda^{TT}_{jilk} \ts, \qquad
(\Lambda^{Tq}_{iabj})^* = \Lambda^{Tq}_{aijb} \ts, \qquad
(\Lambda^{qq}_{abcd})^* = \Lambda^{qq}_{badc} \ts.
\ee
The assumption of time-reversal invariance for this theory before any
potential breaking via vacuum alignment means that the angles $\theta_{TC} =
\theta_{QCD} = 0$ (at tree level) and that all the $\Lambda$'s are
real. Thus, e.g., $\Lambda^{TT}_{ijkl} = \Lambda^{TT}_{jilk}$.

All the four--fermion operators in $\CH'$ are renormalized at the ETC scale.
Throughout this work, we shall assume that the ETC gauge symmetries commute
with electroweak $SU(2)$, but not with weak hypercharge $U(1)$ (indeed, they
must not; see Ref.~\cite{etceekl}). The ETC interactions then take the form,
e.g.,
\be\label{eq:HTT}
\CH'_{TT} =
\left(\ol U_{iL}\gamma^{\mu}U_{jL} + \ol D_{iL}\gamma^{\mu}D_{jL}\right)
\left(\Lambda^U_{ijkl}\ts \ol U_{kR}\gamma_{\mu}U_{lR}
+ \Lambda^D_{ijkl} \ts \ol D_{kR}\gamma_{\mu}D_{lR}\right) \ts.
\ee

Having chosen a standard chiral--perturbative ground state, $|\Omega\rangle$,
vacuum alignment proceeds by minimizing the expectation value of the rotated
Hamiltonian. This is obtained by making the $G_f$ transformation $T_{L,R}
\ra W_{L,R} \ts T_{L,R}$ and $q_{L,R} \ra V_{L,R} \ts q_{L,R}$, where
$W_{L,R} \in SU(2N)_{L,R}$ and $V_{L,R} \in SU(2n)_{L,R}$:
\bea\label{eq:HW}
\CH'(W,V) &=& \CH'_{TT}(W_L,W_R) +  \CH'_{Tq}(W,V) +
\CH'_{qq}(V_L,V_R) \\
&=& \Lambda^{TT}_{ijkl} \ts \ol{T}_{i'L} W_{L\ts i'i}^\dag
\gamma^{\mu}W_{L\ts jj'}T_{j'L} \ts \ol{T}_{k'R} W_{R\ts k'k}^\dag
\gamma^{\mu}W_{R\ts ll'}T_{l'L} + \cdots \ts.\nn
\eea
Since $T$ and $q$ transform according to complex representations of their
respective color groups, the four--fermion condensates in the
$S_f$--invariant $|\Omega\rangle$ have the form
\bea\label{eq:conds}
\langle\Omega|\ol{T}_{iL}\gamma^{\mu}T_{jL}
\ts \ol{T}_{kR}\gamma_{\mu}T_{lR}|\Omega\rangle &=& -\Delta_{TT}
\delta_{il}\delta_{jk} \ts, \nonumber \\
\langle\Omega| \ol{T}_{iL}\gamma^{\mu}q_{aL}
\ts \ol{q}_{bR}\gamma_{\mu}T_{jR} |\Omega\rangle &=&
-\Delta_{Tq}\delta_{ij}\delta_{ab} \ts, \\
\langle\Omega|\ol{q}_{aL}\gamma^{\mu}q_{bL}
\ts \ol{q}_{cR}\gamma_{\mu}q_{dR} |\Omega\rangle &=&
-\Delta_{qq}\delta_{ad}\delta_{bc} \ts. \nonumber 
\eea
The condensates are positive, renormalized at $M_{ETC}$ and, in the
large--$N_{TC}$ and $N_C$ limits, they are given by $\Delta_{TT} \simeq
(\Delta_T(M_{ETC}))^2$, $\Delta_{Tq} \simeq \Delta_T(M_{ETC})$ $\times
\Delta_q(M_{ETC})$, and $\Delta_{qq} \simeq (\Delta_q(M_{ETC}))^2$. In
walking technicolor, $\Delta_T(M_{ETC}) \simeq (M_{ETC}/\Lambda_{TC}) \ts
\Delta_T(\Lambda_{TC})$ $= 10^2$--$10^3 \times \Delta_T(\Lambda_{TC})$. In
QCD, however, $\Delta_q(M_{ETC}) \simeq
(\log(M_{ETC}/\Lambda_{QCD}))^{\gamma_m} \ts \Delta_q(\Lambda_{ETC}) \simeq
\Delta_q(\Lambda_{QCD})$, where $\gamma_m \simeq 2\alpha_C/\pi$ for
$SU(3)_C$~\cite{kdl}. Thus, the ratio
\be\label{eq:ratio}
r = {\Delta_{Tq}(M_{ETC}) \over{\Delta_{TT}(M_{ETC})}} \simeq
{\Delta_{qq}(M_{ETC}) \over{\Delta_{Tq}(M_{ETC})}}
\ee
is at most $10^{-10}$. This is 10--100 times smaller than in a technicolor
theory in which the coupling does not walk.

With these condensates, the vacuum energy is a function only of $W = W_L \ts
W_R^\dag$ and $V = V_L \ts V_R^\dag$, elements of the coset space $G_f/S_f$:
\bea\label{eq:vacE}
& &E(W,V) = E_{TT}(W) + E_{Tq}(W,V) + E_{qq}(V) \\
& & \ts\ts = -\Lambda^{TT}_{ijkl} \ts W_{jk} \ts W^\dag_{li} \ts \Delta_{TT}
       -\left(\Lambda^{Tq}_{iabj} \ts V_{ab} \ts W^\dag_{ji} + {\rm c.c.}
         \right) \Delta_{Tq} 
       -\Lambda^{qq}_{abcd} \ts V_{bc} \ts V^\dag_{da} \ts \Delta_{qq}
       \ts.\nn 
\eea
Note that time--reversal invariance of the unrotated Hamiltonian $\CH'$
implies that $E(W,V) = E(W^*,V^*)$. Hence, spontaneous CP--violation occurs
if the solutions $W_0$, $V_0$ to the minimization problem are complex.

Following Ref.~\cite{elp}, we define technifermion current mass matrices
renormalized at the ETC scale as follows:\footnote{These definitions differ
  from those in Ref.~\cite{elp} by a common vectorial transformation on the
  left and right--handed fields. This affects none of our discussion.}
\bea\label{eq:Tmass}
M_{Tij} \ts \Delta_T(M_{ETC}) &=& - \left(W^\dag_{ik} {\partial E \over
    {\partial W^\dag_{jk}}} \right)_{W_0,V_0} \nn\\
 &=& W^\dag_{0ik} \left(\Lambda^{TT}_{klmj} W_{0lm} \Delta_{TT} 
   +   \Lambda^{Tq}_{kabj} V_{0ab} \Delta_{Tq} \right) \nn\\
 &=& W^\dag_{0ik} \ts \Lambda^{TT}_{klmj} \ts W_{0lm} \ts \Delta_{TT} \left(1
   + \CO(r)\right) \ts.
\eea 
For quarks,
\bea\label{eq:qmass}
M_{qab} \ts \Delta_q(M_{ETC}) &=& - \left(V^\dag_{ac} {\partial E \over
    {\partial V^\dag_{bc}}} \right)_{W_0,V_0} \nn\\
 &=& V^\dag_{0ac} \left(\Lambda^{Tq}_{cijb} W_{0ij} \Delta_{Tq} 
   +   \Lambda^{qq}_{cdeb} V_{0de} \Delta_{qq} \right) \nn\\
 &=& V^\dag_{0ac} \ts \Lambda^{Tq}_{cijb} \ts W_{0ij} \ts \Delta_{Tq}
\left(1 + \CO(r)\right) \ts.
\eea 
Nuyts' theorem generalized to technicolor~\cite{nuyts,elp} states that, as a
consequence of extremizing the energy, the imaginary parts of these matrices
are proportional to the identity:
\bea\label{eq:nuyt}
& & \left(M_T - M^\dag_T \right)\Delta_T(M_{ETC})  = i\nu_T 1_{2N} \ts, \nn\\
& & \left(M_q - M^\dag_q \right)\Delta_q(M_{ETC})  = i\nu_q 1_{2n} \ts.
\eea
The parameters $\nu_T$ and $\nu_q$ are Lagrange multipliers associated with
the unimodularity constraints on $W_0$ and $V_0$, respectively.  These
equations imply that $M_T$ and $M_q$ are each diagonalized by a (different)
single special unitary transformations. Taking the trace of both sides of
Eqs.~(\ref{eq:nuyt}) and using Eqs.~(\ref{eq:Tmass},\ref{eq:qmass}) gives
\bea\label{eq:trace}
2iN\nu_T &\equiv& {\rm Tr}\left(M_T - M^\dag_T \right)\Delta_T(M_{ETC}) \nn\\
&=& -{\rm Tr}\left(M_q - M^\dag_q \right)\Delta_q(M_{ETC}) \equiv -2in\nu_q
\nn\\
&=& 2i \Lambda^{Tq}_{kabj} \ts
{\rm Im} \left(W^*_{0kj} V_{0ab}\right) \Delta_{Tq}
\ts.
\eea
This relation between $\nu_T$ and $\nu_q$ requires that $SU(N_{TC})$ and
$SU(3)_C$ are embedded in a simple ETC group, so that $\theta_{TC} =
\theta_{QCD}$.

Strong CP--violation occurs if $\nu_{T,q} \neq 0$. The angle $\ol \theta_q$
characterizing this for quarks' is defined by (for $\theta_{QCD} =
0$)~\footnote{See, e.g., Ref.~\cite{baluni} for the relation between
  $\theta_q$ and $\nu_q$ for the case of three light quarks.}
\be\label{eq:thbar}
\ol \theta_q = \arg\ts\det \left(M_q\right) \Delta_q =
\arg\ts\det\left(\CM_q \right) \Delta_{q} \ts,
\ee
where, up to $\Lambda^{qq}$ terms of relative order $r$,
\be\label{eq:cmq}
\CM_{qab} \ts \Delta_q \equiv (V_0 M_q)_{ab} \ts \Delta_q =
\Lambda^{Tq}_{aijb} \ts W_{0ij} \ts \Delta_{Tq}
\ee
is the primordial quark mass matrix, i.e., the one {\it before} vacuum
alignment in the quark sector. We see that strong CP--violation arises from a
conflict between mass terms and a chiral symmetry constraint on the alignment
matrix. This is what Dashen and Nuyts showed for quarks in QCD and what we
found in Ref.~\cite{elp} for extended technicolor. In a world with just one
type of fermion, say $T_{i L,R}$, with explicit flavor symmetry breaking due
to gauge interactions {\it alone}, $M_T = M_T^\dag$ and there is no strong
CP--violation even if the aligning matrix $W_0$ is complex and CP symmetry
is spontaneously broken,

Suppose we found $\ol \theta_q = 0$ up to the $\Lambda^{qq}$ terms of order
$r$. Are there larger contributions to $\ol \theta_q$?  The first to worry
about are two--loop ETC contributions to $\CM_q$. There are two types of
these: those with one technifermion dynamical mass insertion and those with
three. The first are proportional to a single power of $W_0$ and, because the
$\Lambda^{TT}$'s are real, it is plausible that they will not change $\ol
\theta_q$. This must be checked in specific models. The three--insertion
graphs involve two $W_0$'s and one $W^\dag_0$ convoluted with
$\Lambda^{TT}$'s and these are more likely to contribute to $\ol \theta_q$.
Apart from any $g^2_{ETC}/16\pi^2$ suppression these graphs may have, they
are of relative order $\Delta^2_T(M_{ETC})/M^6_{ETC} \simle
(\Lambda_{TC}/M_{ETC})^4 \simle 10^{-10}$ in a walking technicolor theory. We
tentatively conclude that the $\ol \theta_q$ defined in Eq.~(\ref{eq:thbar})
is a reliable measure of strong CP--violation in extended technicolor
models. We need only know $W_0$ and the $\Lambda^{Tq}$ to determine it.

Our strategy for vacuum alignment, which we carry out numerically, is the
following: Because $r$ is small, we first minimize $E_{TT}$ to determine
$W_0$. If we wish to determine $W_{0L}$ and $W_{0R}$ separately, we make
vectorial transformations on $T_{L,R}$ that diagonalize $M_{Tij}$.  Physical
results such as technipion and quark masses are unchanged even if we use, for
example, $W_{0L} = W_0$. The results of technifermion alignment are presented
in the next section.

Once $W_0$ is determined, it is inserted as a set of parameters into $E_{Tq}$
and this is minimized as a function of $V$. If there are several degenerate
solutions $W_0$ minimizing $E_{TT}$, one should choose the one giving the
deepest minimum $E_{Tq}(W_0,V_0)$.  When $V_0$ is known, the matrices
$V_{0L}$, $V_{0R}$ are determined by diagonalizing the matrix $M_q$ in
Eq.~(\ref{eq:qmass}). The quark CKM matrix is then obtained from $V_{0L}$.

Finally, holding $V_0$ fixed, one can refine $W_0$ by reminimizing $E_{TT} +
E_{Tq}$ as a function of $W$. This will induce corrections of $\CO(r)$ in
$W_0$ and $\ol \theta_q$. There is no point in refining $V_0$ by minimizing
the full energy including $E_{qq}$. However, note that the rotated
$\CH'_{qq}(V_0)$ may contain sources of quark CP--violation not contained in
the CKM matrix~\cite{elp}. These studies of CP--violation in the quark sector
will be presented in our next paper.

We are concerned in this paper with vacuum alignment in the technifermion
sector, and we turn to this now. We will allow only models in which alignment
conserves electric charge, i.e., does not induce $\ol U_i D_j$
condensates. Then, the matrix minimizing $E_{TT}(W)$ must be block--diagonal,
\be\label{eq:Wblock}
W_0 = \left(\ba{cc} W^U_0 & 0 \\ 0 & W^D_0 \ea\right) \ts,
\ee
where $W^U_0$, $W^D_0$ are $U(N)$ matrices satisfying $\det(W^U_0)
\det(W^D_0)=1$. The phase indeterminacy of the individual $U(N)$ matrices
corresponds to the electroweak $T_3$ symmetry. Thus, for admissible models,
we can minimize $E_{TT}$ in the subspace of block--diagonal matrices. Using
Eq.~(\ref{eq:HTT}), the vacuum energy takes the form
\bea\label{Eblock}
E_{TT}(W^{U},W^{D}) &=&
-(\Lambda^{U}_{ijkl} \ts W^{U}_{jk}\ts W^{U\dag}_{li}
+ \Lambda^{D}_{ijkl} \ts W^{D}_{jk}\ts W^{D\dag}_{li})\Delta_{TT} \nn \\
&\equiv& E_U(W^U) + E_D(W^D) \ts.
\eea
Since this expression is bilinear in $W^{U,D} \ts W^{U,D\ts\dag}$, without
loss of generality we can determine $W_0$ by separately minimizing $E_U$ and
$E_D$ in the space of $SU(N)$ matrices. We do this in the next section,
taking care to ensure that no Goldstone bosons remain massless other than the
three associated with electroweak $SU(2)$ symmetry. This means that
$\Lambda^{U,D}_{ijkl}$ must be chosen so that there are no massless $SU(N)
\otimes SU(N)$ Goldstone boson in either $U$ or $D$ sector.

We close this section with some remarks on calculating the pseudoGoldstone
boson (technipion) mass spectrum in these technicolor models. In the standard
chiral--perturbative ground state, $|\Omega\rangle$, the spontaneously broken
symmetries are formally generated by the $G_f/S_f$ charges
\be\label{eq:qfive} Q^A_5 = \half \int d^3x
\left(T^\dag_{R} \lambda_{A} T_{R} -
  T^\dag_{L} \lambda_{A} T_{L}\right) \ts,
\ee
Here, $\lambda_A$ are the $4N^2-1$ Gell-Mann matrices of $SU(2N)$. To first
order in the chiral perturbation $\CH'_{TT}$, technipion masses are given by
Dashen's formula~\cite{rfd},
\be\label{eq:dashen}
F^2_T M^2_{\pi \ts AB} = i^2 \left\langle\Omega\left| \left[Q^A_5, 
\left[Q^B_5,\CH'_{TT}(W_{0L}, W_{0R})\right]\right]
\right|\Omega\right\rangle \ts,
\ee
where $F_T = 246\,\gev/\sqrt{N}$ is the technipion decay constant and
$\CH'_{TT}$ is given in Eq.~(\ref{eq:HW}). As noted above, we can determine
$W_{0L}$ and $W_{0R}$ by diagonalizing the technifermion current--mass
matrix, $M_T$. However, since $M^2_{\pi \ts AB}$ is invariant under vectorial
transformations of the technifermion fields, it is simpler to compute it using
$W_{0L} = W_0$ and $W_{0R} = 1$. The result is
\bea\label{eq:tpimass}
& &F^2_T M^2_{\pi \ts AB} = \half \Lambda_{ijkl} \Biggl[
\left(\left\{ \lambda_A , \lambda_B\right\}W_0^\dag\right)_{li} \ts W_{0jk} +
\left(W_0\left\{\lambda_A , \lambda_B\right\}\right)_{jk} \ts W_{0li}^\dag
\nn\\
& &\qquad -2 \left(\lambda_A W_0^\dag\right)_{li} \ts
\left(W_0 \lambda_B \right)_{jk} 
-2 \left(\lambda_B W_0^\dag\right)_{li} \ts
\left(W_0 \lambda_A \right)_{jk}\ts \Biggr]\Delta_{TT} \ts.
\eea
Note that, because the vector charges annihilate the standard vacuum, the
axial charges in Eq.~(\ref{eq:tpimass}) may be replaced by left--handed or
right--handed charges or by any linear combination that is not purely a
vector charge.

In using Eq.~(\ref{eq:tpimass}) in these models, we have seen examples in
which a technipion's mass vanishes without there being a corresponding
conserved chiral charge, i.e., a linear combination of the $Q^A_R$ and
$Q^A_L$ which commutes with $\CH'_{TT}(W_0)$. A two--technidoublet, $SU(4)
\otimes SU(4)$ example is provided by the following set of $\Lambda$'s (whose
scale is arbitrary):
\bea\label{eq:modela}
\Lambda^U_{1111} &=& \Lambda^D_{1111} = \Lambda^U_{2222} =
  \Lambda^D_{2222} = 1 \nn\\
\Lambda^U_{1112} &=& \Lambda^U_{1121} = \Lambda^D_{1112} =
  \Lambda^D_{2221} = \half \nn\\
\Lambda^U_{1211} &=& \Lambda^U_{2111} = \half \ts.
\eea
In addition to the three electroweak Goldstone bosons coupling to
$$
\half \int d^3x \ts \sum_{i=1}^2 \ts \left(U_i^\dag, D_i\right) \gamma_5
  \tau_a \left(\ba{c}U_i \\D_i \ea\right) \ts,
$$
there is a fourth one associated with the $W_0$--rotation of the axial charge
$$
\half \int d^3x \left(D_1^\dag \gamma_5 D_1 - D_2^\dag \gamma_5 D_2 \right)
\ts.
$$
However, the divergence of its current is manifestly of first order in
$\CH'_{TT}(W)$.

This extra massless technipion is at first surprising when one recalls that
Dashen proved that a zero eigenvalue of the Goldstone boson mass--squared
matrix implies that the corresponding current is conserved~\cite{rfd}.
Furthermore, in QCD we have become used to a conserved current being
associated with a massless Goldstone boson.  There, the symmetry that leaves
the boson massless is manifest in the mass matrix $M_q$ of $\CH'_q = \ol q_L
M_q \ts q_R + {\rm h.c.}$. The resolution of this puzzle is that Dashen's
proof applies to the matrix elements of double commutators in the {\it exact}
ground state, $|{\rm vac}\rangle$, of the {\it full} Hamiltonian $H = \int
(\CH_0 + \CH'(W))$. The matrix in Eq.~(\ref{eq:tpimass}) is calculated in the
perturbative ground state, $|\Omega\rangle$, which is the limit of $|{\rm
vac}\rangle$ as $\CH'(W) \ra 0$. Consequently, all that can be proved for a
``massless'' Goldstone boson at the perturbative level at which we work is
that {\it all} matrix elements of the divergence of the corresponding current
must be of {\it second} order in $\CH'$. We emphasize that, although the
masslessness of this technipion may be an approximation, it is important
phenomenologically. Corrections to its mass are likely to be so small that it
is already ruled out experimentally.

\section*{3. Results From the Technifermion Sector}

The vacuum energy in the $U$ and $D$--technifermion sectors has the form
\bea\label{eq:EW}
E(W) &=& - \sum_{ijkl=1}^N \ts \Lambda_{ijkl} \ts W_{jk} W^\dag_{li} \ts
\Delta_{TT} \nn\\
&\equiv&  - \sum_{ijkl=1}^N \ts \Lambda_{ijkl} \ts |W_{jk}|
|W_{il}| \exp[i(\phi_{jk} - \phi_{il})] \ts \Delta_{TT}\ts,
\eea
where $W = W_U$ or $W_D \in SU(N)$, $\Lambda_{ijkl} = \Lambda^*_{ijkl} =
\Lambda_{jilk}$, and $\phi_{jk} = \arg(W_{jk})$. We remind the reader that we
always choose the $\Lambda_{ijkl}$ so that there is no $SU(N) \otimes SU(N)$
Goldstone boson in either $U$ or $D$ sector. Note that, if $W_0$ minimizes
$E(W)$, then so do the matrices $Z_N^{2m} \ts W_0 = \exp(2im\pi/N)W_0$, $m =
1,2,\dots,N$, and their complex conjugates. This degeneracy may be lifted by
the quark--technifermion interaction $\CH'_{Tq}$.

It is especially convenient to parameterize $W$ in the form
\be\label{eq:Wform}
W = D_L K D_R \ts.
\ee
Here, $D_{L,R}$ are diagonal unimodular matrices, each depending on $N-1$
phases:
\be\label{eq:Dform}
D_{L,R} = {\rm diag} \ts \left[\exp(i \chi_{L,R \ts 1}), \exp(i \chi_{L,R\ts
2}), \dots, \exp(-i (\chi_{L,R \ts 1} + \cdots + \chi_{L,R \ts N-1})) 
\right] \ts,
\ee
and $K$ is an $(N-1)^2$--parameter CKM matrix which we write in the standard
Harari--Leurer form~\cite{harari}. This matrix depends on $\half N(N-1)$
angles $\theta_{ij}$, $1\le i < j \le N$, and $\half (N-1)(N-2)$ phases
$\delta_{ij}$ , $1\le i < j-1 \le N-1$.

We have discovered several remarkable properties of the matrices $W_0$ which
minimize $E(W)$. They have to do with the fact that the coefficient
$\Lambda_{ijkl}$ tends to align the phases $\phi_{jk}$ and $\phi_{il}$: if
$\Lambda_{ijkl} > 0$, its contribution to $E(W)$ is minimized if the
phases can be equal; if $\Lambda_{ijkl} < 0$, the phases want to differ by
$\pi$. Of course, because not all the $N^2$ phases $\phi_{jk}$ are
independent, unitarity can frustrate phase alignment. If the nonzero
$\Lambda_{ijkl}$ link {\it all} the $\phi_{jk}$ together, then all of them
will be equal, mod~$\pi$, {\it if that can be consistent with
unitarity}. Unimodularity then requires all $\phi_{jk} = 2m\pi/N$, mod~$\pi$,
with $m=1,2,\dots,N$. We call this ``complete phase alignment'' and we say
that the phases are ``rational''.

Rational phases may also occur when the nonzero $\Lambda_{ijkl}$ link some,
but not all, of the phases. If it is allowed by unitarity, we have found that
the phases are multiples of $\pi/N'$ (modulo the $Z_N$ phase $2m\pi/N$) for
one or more values of $N'$ between~1 and~$N$. This case of partial phase
alignment is very rich, with many possibilities and, sometimes, degenerate
minima whose $W_0$'s are not unitarily equivalent nor related by conjugation
or a $Z_N$ factor. Its implications for quark CP--violation will be studied
in our next paper.

A necessary condition for phase alignment is that the CKM matrix $K$ is
real. The reason for this is seen by looking at a typical complex term in
$K$ for the 3$\times$3 case, e.g., $s_{12} \ts s_{23} - c_{12} \ts c_{23}
\ts s_{13} \exp{(i\delta_{13})}$, where $s_{12} = \sin\theta_{12}$, etc. The
mixing angles $\theta_{ij}$ are determined by the $\Lambda$'s that are
dominant in minimizing the energy and by unitarity. Then, the overall phase
of this term will be a random irrational number unless $\delta_{13} = 0$ or
$\pi$ or one of the $\theta_{ij} = 0$. If $K$ is complex, it contains more
random phases than can be made rational by choices of phases in $D_L$ and
$D_R$, and so the $\phi_{jk}$ will be randomly irrational. Note that the case
$N=2$ is special because $K$ is always real. In that case, all phases in
$W_0$ are $0$ or $\pi/2$, mod~$\pi$.

Suppose that completely or partially--aligned rational phases occur for some
set of $\Lambda$'s. Then we find that the nonzero $\Lambda$'s may be varied
over an appreciable range with no change whatever in the phases. Ultimately,
a large enough excursion in the $\Lambda$'s will make it impossible to
maintain unitarity with aligned phases and, at certain critical values of the
$\Lambda$'s, they change continuously from rational to irrational (or, in the
$SU(2)$ case, discontinuously from one rational set to another). A
rational--to--irrational phase transition may also occur if vanishing
$\Lambda$'s are made nonzero. By further varying the $\Lambda$'s, another,
possibly inequivalent, set of rational phases may characterize the matrix
$W_0$. Thus, the minima of $E(W)$ as one varies the $\Lambda$'s are islands
of rational aligned phases in a sea of irrational ones.

A Goldstone boson appears whenever a transition occurs between different
types of phases. As the critical $\Lambda$'s are approached, one of the
$M_\pi^2$ decreases to zero and then increases again once the boundary is
passed. What is happening is this: As the transition is approached, the
ground states for a set of rational phases are becoming less stable and a
technipion's $M_\pi^2$ is diving through zero to negative values. At the same
time, the ground states for a nearby set of irrational phases are becoming
more stable and the corresponding $M_\pi^2$ is increasing from negative to
positive values. The two types of phases coexist at the rational island
shore, giving rise to infinitely many degenerate minima that are
characterized by an indeterminacy in the phases of $D_{L,R}$ and $K$. Hence,
$M_\pi^2 = 0$ (to first order) there. This is another situation in which the
massless state's chiral charge does not commute with $\CH'_{TT}$.

This phenomenon may be important. The appearance of an exceptionally light
technipion is not uncommon because typical rational--phase
$\Lambda$--parameters often are not far from critical ones. In
Ref.~\cite{elw} we observed that, because the number of technidoublets in
typical TC2 models is large, $N \sim 10$, the technihadron scale is low and
technipion masses may be as light as 100~GeV. Now we see that some
technipions may be even lighter than nominally expected from the
$\Lambda$'s. In a specific model, this may be a major prediction or it may be
a show--stopper.

Finally, another interesting property of the rational--phase minima is that
the coefficients $\tilde{\Lambda}_{ijkl} = \sum_{i'j'} \ts \Lambda_{i'j'kl}
\ts W^\dag_{0\ts ii'} \ts W_{0\ts j'j}$ in the rotated Hamiltonian
\be\label{eq:Hrot}
\CH'_{TT}(W_0) = \tilde{\Lambda}_{ijkl} \ts \ol T_{Li} \ts \gamma^\mu \ts
T_{Lj} \ts \ol T_{Rk} \ts \gamma_\mu \ts T_{Rl} \ts, 
\ee
also have rational phases. This follows directly from the fact that nonzero
$\Lambda$'s align phases. If the phases are rational and $\Lambda_{i'j'kl}
\neq 0$, then the CKM matrix $K$ is real and $\phi_{j'k} - \phi_{i'l} =
\chi_{Lj'} - \chi_{Rk} - \chi_{Li'} + \chi_{Rl} = 0$ (mod~$\pi$).  The phase
of an individual term in the sum for $\tilde \Lambda_{ijkl}$ is then
$\phi_{j'j} - \phi_{i'i} = \chi_{Lj'} - \chi_{Rj} - \chi_{Li'} + \chi_{Ri} =
\chi_{Rk} - \chi_{Rj} - \chi_{Rl} + \chi_{Ri}$ (mod~$\pi$), a rational phase
which is the same for all terms in the sum over $i', j'$.

One example of these phenomena is provided by an $SU(3)$ model in which
the nonzero $\Lambda$'s are:
\bea\label{eq:modelb}
&&\Lambda_{1111} = \Lambda_{1221} = \Lambda_{2112} =
  \Lambda_{1212} = \Lambda_{2121} = 1.0 \nn\\
&&\Lambda_{1122} = 1.5 \ts, \quad \Lambda_{1133} = 1.4 \nn\\
&&  \Lambda_{1331} = \Lambda_{3113} = 1.6, \quad
\Lambda_{1313} = \Lambda_{3131} = 1.8 \nn\\
&&\Lambda_{1222} = \Lambda_{2122} = \Lambda_{2212} = \Lambda_{2221} =
0.50-1.1 \ts.
\eea
These want to align $\phi_{11} = \phi_{22} = \phi_{33} = \phi_{12} =
\phi_{21}$ and $\phi_{13} = \phi_{31}$. Phases $\phi_{23}$ and $\phi_{32}$
are not linked by the $\Lambda$'s. The effect of varying $\Lambda_{1222}$
from 0.5 to 1.1 is illustrated in Fig.~1. The phases start out aligned and
rational, indeed, $W = \exp{(2i\pi/3)} \times 1$, and the vacuum energy (in
units of $\Delta_{TT}$) remains constant at $-6.20$. At $\Lambda_{1222}
\simeq 0.725$, it becomes energetically favorable for $W$ to become
nondiagonal. The phases are still aligned and rational, equal to $2\pi/3$
(mod~$\pi/2$), and a technipion mass becomes zero here. Now the energy
decreases as $\Lambda_{1222}$ is increased. At $\Lambda_{1222} \simeq 1.015$,
a second transition occurs in which rational phases are no longer possible,
and a different technipion's mass goes through zero. At $\Lambda_{1222}
\simeq 1.045$, a transition occurs back to rational phases, all equal to
$\pi/3$ (mod~$\pi)$, and the same technipion's mass vanishes
again. Throughout this variation of $\Lambda_{1222}$, the other six
technipion squared masses remain fairly constant with values between~5
and~15. Thus, in this example, the two technipion masses shown in Fig.~1 are
always quite light.

\section*{4. Summary and Outlook}

We have numerically studied vacuum alignment in a class of theories in which
electroweak and flavor symmetries are dynamically broken by gauge
interactions alone. To make these intial studies tractable, we considered
extended technicolor with $N$ doublets of a single type of technifermion,
$T_{iL,R}$, transforming according to a complex representation of
$SU(N_{TC})$ but as $SU(3)_C$ singlets. These were coupled by ETC to $n$
quark doublets, $q_{aL,R}$. In the absence of an explicit model for ETC, we
assumed its broken gauge interactions could produce any desired four--fermion
interaction of the form $\CH'$ in Eq.~(\ref{eq:Hetc}). As usual, we assumed
that ETC commutes with electroweak $SU(2)$, but not $SU(N_{TC}) \otimes
SU(3)_C \otimes U(1)$~\cite{etceekl}. We also assumed, quite naturally, that
ETC breaking preserves CP--invariance so that the $\Lambda$ parameters in
$\CH'$ are real.

We focussed on the technifermion sector in this paper. This restriction
determines the vacuum--aligning matrix $W_0$ of technifermions up to tiny,
but potentially important corrections of order $\langle\ol q q \rangle_{ETC}
/ \langle\ol T T \rangle_{ETC} \sim 10^{-10}$. The problem is then simplified
both numerically and analytically to minimizing the vacuum energy $E_{TT}(W)$
in the subspace of up and down--block diagonal $W$--matrices which conserve
electric charge. We need then only study the alignment problem in a single
charge sector. To ensure that no technipions remain massless other than the
three associated with electroweak symmetry, the ETC parameters
$\Lambda^{U,D}_{ijkl}$ in $\CH'_{TT}$ must be chosen so that there are no
$SU(N) \otimes SU(N)$ Goldstone boson in either $U$ or $D$ sector.

We found several interesting features of vacuum alignment:

\begin{enumerate}

\item A technipion mass may vanish to first order in the symmetry breaking
perturbation even if its chiral charge does not commute with
$\CH'_{TT}(W_0)$. This differs from what happens in QCD and
$\Sigma$--model--like effective Lagrangians where the symmetries of the
perturbation are manifest. The reason for this difference is the
four--fermion nature of $\CH'_{TT}$ and the symmetries of the zeroth--order
ground state $|\Omega\rangle$.

\item The real parameter $\Lambda^{U,D}_{ijkl}$ links the $W^{U,D}$ phases
$\phi^{U,D}_{jk}$ and $\phi^{U,D}_{il}$. If allowed by unitarity of $W$,
these phases are then equal or differ by $\pi$. If there is complete phase
alignment, all phases are equal to integer multiples of $2\pi/N$,
mod~$\pi$. If only partially aligned, the phases are integer multiples of
$\pi/N'$ for one or more $N' \le N$. If phase alignment is inconsistent with
unitarity, the phases are irrational multiples of $\pi$.

\item Rational phase sets are natural in the sense that they remain unchanged
for a finite range of $\Lambda$ parameters. In $\Lambda$--space, the rational
phase solutions to vacuum alignment form discrete islands in a sea of
irrational phase solutions.

\item A massless (to first order) Goldstone boson appears when the
$\Lambda$'s take on critical values defining the boundary between rational
and irrational phases. Thus, exceptionally light technipions are not at all
uncommon and are a new phenomenological consequence of vacuum alignment.

\end{enumerate}

Vacuum alignment in the quark sector and the central issue of quark
CP--violation will be addressed in a subsequent paper. It is obvious from
Eq.~(\ref{eq:trace}) that irrational phases in the technifermion matrix $W_0$
will induce strong CP--violation for quarks: $\nu_q \neq 0$. It is therefore
fortunate that rational phases occur naturally. They may permit a dynamical
theory of quark flavor in which only weak CP--violation occurs and in which
there is no axion.

\section*{Acknowledgements}

We thank Sekhar Chivukula for his helpful comments on the manuscript.
The research of KL and TR is supported in part by the Department of Energy
under Grant~No.~DE--FG02--91ER40676. The research of EE is supported by the
Fermi National Accelerator Laboratory, which is operated by Universities
Research Association, Inc., under Contract~No.~DE--AC02--76CHO3000.

\vfil\eject

\begin{figure}[tb]
\vbox to 12cm{
\vfill
\includegraphics{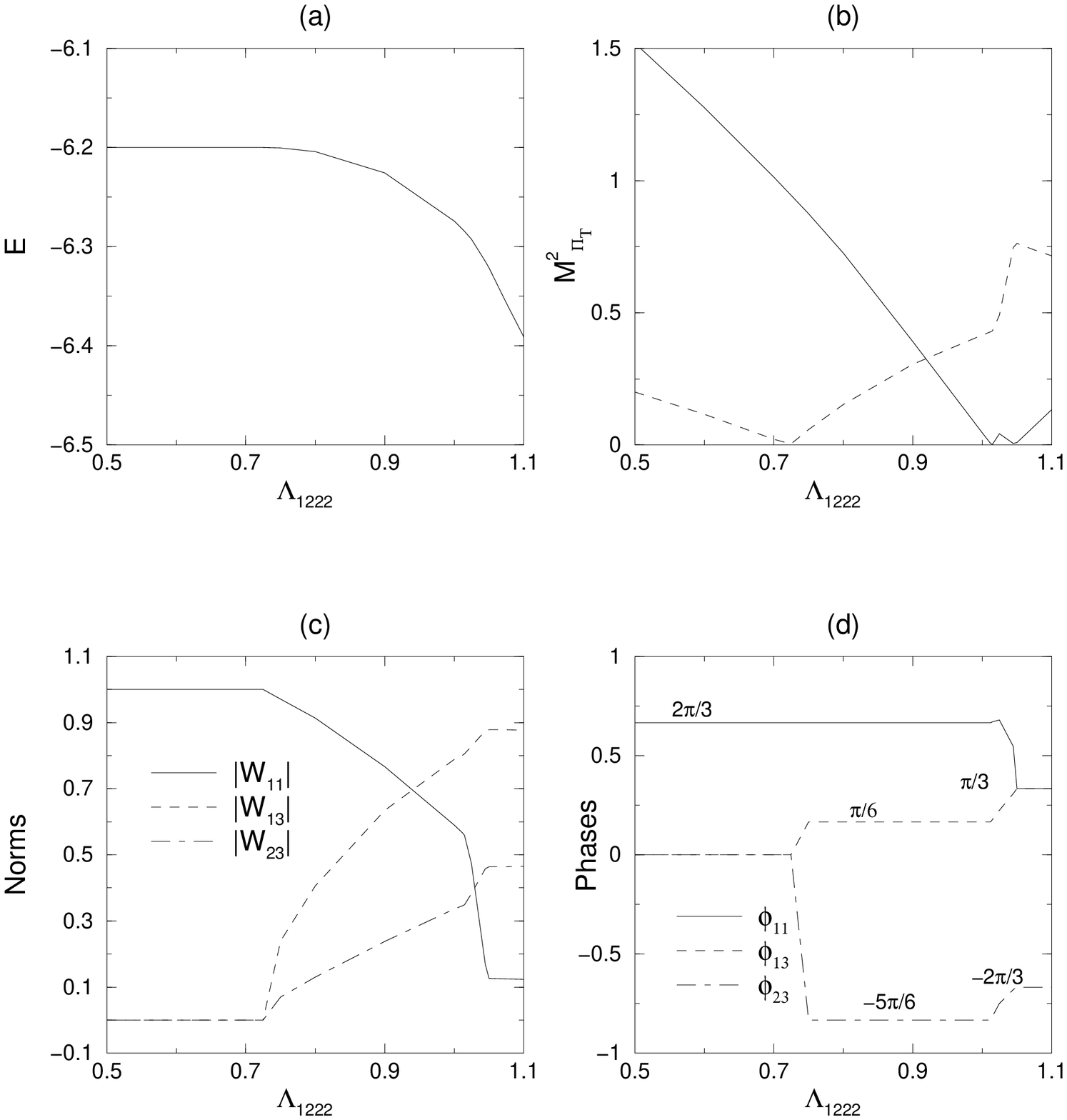}
\vfill
}
\caption{Phase alignment in a model with $N=3$ as a function $\Lambda_{1222}$;
other $\Lambda$--parameters are fixed in Eq.~(25). (a) The vacuum
energy, $E(W)$ (arbitrary units); (b) the squared mass of two of the
technipions;  (c), (d) the magnitudes and phases of
$W_{11}$, $W_{13}$ and $W_{23}$.}

\label{fig:one}
\end{figure}


\begin{thebibliography}{99}
%
\bibitem{tc} S.~Weinberg, Phys.~Rev.~{\bf D19}, 1277 (1979);\\
L.~Susskind, Phys.~Rev.~{\bf D20}, 2619 (1979).
%
%
\bibitem{etceekl} E.~Eichten and K.~Lane, Phys.~Lett.~{\bf B90}, 125 (1980).
%
%
\bibitem{etcsd}S.~Dimopoulos and L.~Susskind, Nucl.~Phys.~{\bf B155}, 237
(1979).
%
%
\bibitem{CPreview} See, e.g., R.~D.~Peccei, {\it QCD, Strong CP and Axions},
  hep-ph/9606475.
%
%
\bibitem{rfd}R.~F.~Dashen, Phys.~Rev.~{\bf D3}, 1879 (1971).
%
%
\bibitem{nuyts} J.~Nuyts, Phys.~Rev.~Lett.~{\bf 26}, 1604 (1971). Nuyts'
  theorem applies even if $M_q \neq M_q^*$; it requires only that the
  bilinear condensates are proportional to a unitary matrix.
%
%
\bibitem{elp}E.~Eichten, K.~Lane and J.~Preskill, Phys.~Rev.~Lett.~{\bf 45},
  225 (1980); \\
K.~Lane, Physica Scripta {\bf 23}, 1005 (1981).
%
%
\bibitem{tctwohill}C.~T.~Hill, Phys.~Lett.~{\bf 345B}, 483 (1995).
%
%
\bibitem{tctwoklee}K.~Lane and E.~Eichten, Phys.~Lett.~{\bf B352}, 382
(1995);\\
K.~Lane, Phys.~Rev.~{\bf D54}, 2204 (1996);\\
K.~Lane, Phys.~Lett.~{\bf B433}, 96 (1998).
%
%
\bibitem{iso} T.~Appelquist, et al., Phys.~Rev.~Lett~{\bf 53}, 1523 (1984);
  Phys.~Rev.~{\bf D31}, 1676 (1985);\\
R.~S.~Chivukula, {\it Phys.~Rev.~Lett.}~{\bf 61}, 2657 (1988).
%
%
\bibitem{zbbth}R.~S.~Chivukula, S.~B.~Selipsky, and E.~H.~Simmons,
Phys.~Rev.~Lett.~{\bf 69}, 575 (1992);\\
R.~S.~Chivukula, E.~H.~Simmons, and J.~Terning,
Phys.~Lett.~{\bf B331}, 383, (1994), and references therein.
%
%
\bibitem{cthill}We thank C.~T.~Hill for emphasizing this point to us.
%
\bibitem{elw} K.~Lane and E.~Eichten, Phys. Lett. {\bf B222}, 274 (1989);\\
E.Eichten and K.~Lane, Phys.~Lett.~{\bf B388}, 803 (1996),
hep-ph/9607213;\\
E.~Eichten, K.~Lane and J.~Womersley), Phys.~Lett.~{\bf B405}, 305 (1997),
hep-ph/9704455.
%
%
\bibitem{bbvac}B.~Balaji, Phys.~Lett.~{\bf B393}, 89 (1997), hep-ph/9610446.
%
%
\bibitem{jpmp}J.~Preskill, Nucl.~Phys.~{\bf B177}, 21 (1981); \\
M.~E.~Peskin, Nucl.~Phys.~{\bf B175}, 197 (1980).
%
%
\bibitem{wtc}B.~Holdom, Phys.~Rev.~{\bf D24}, 1441 (1981);
Phys.~Lett.~{\bf 150B}, 301 (1985);\\
T.~Appelquist, D.~Karabali and L.~C.~R. Wijewardhana,
Phys.~Rev.~Lett. {\bf 57}, 957 (1986);\\
T.~Appelquist and L.~C.~R.~Wijewardhana, Phys.~Rev.~{\bf D36}, 568
(1987);\\
K.~Yamawaki, M.~Bando and K.~Matumoto, Phys.~Rev.~Lett.~{\bf 56}, 1335
(1986);\\
T.~Akiba and T.~Yanagida, Phys.~Lett.~{\bf 169B}, 432 (1986).
%
%
%
%
\bibitem{kdl} K.~Lane, Phys.~Rev.~{\bf D10}, 2605 (1974).
%
\bibitem{harari} H.~Harari and M.~Leurer, Phys.~Lett.~{\bf B181}, 123
  (1986);\\
  Review of Particle Properties, C.~Caso, et al., Eur.~Phys.~J.~{\bf C3}, 1
  (1998).
%
%
\bibitem{baluni} V.~Baluni, Phys.~Rev.~{\bf D19}, 2227 (1979).

\end{thebibliography}
\end{document}